\documentclass[conference, english, 10pt]{IEEEtran}
\IEEEoverridecommandlockouts

\usepackage{babel}
\usepackage[babel]{microtype}
\usepackage[babel]{csquotes}

\usepackage[T1]{fontenc}
\usepackage{textcomp}

\usepackage[svgnames]{xcolor}
\usepackage{graphicx}
\usepackage{tikz}
\usepackage{pgfplots}
\usepackage{subfigure}

\usepackage{tabularx}
\usepackage{booktabs}

\usepackage{amsmath}
\usepackage{amsfonts}
\usepackage{amssymb}
\usepackage{amsthm}
\usepackage{algorithm,algpseudocode}
\usepackage{bm}
\usepackage{siunitx}
\usepackage{mathtools}
\usepackage{dsfont}

\usepackage{xparse}

\usepackage[backend=biber, style=ieee, url=false, isbn=false, dashed=false, doi=true]{biblatex}

\bibliography{ref.bib}

\usepackage{hyperref}
\hypersetup{
	pdfauthor={},
	pdftitle={},
	bookmarksnumbered=true,
	colorlinks=true,
	allcolors=.,
	urlcolor=MidnightBlue,
}

\usepackage[acronyms,nomain,xindy,nowarn]{glossaries}
\makeglossaries
\newacronym{ao}{AO}{alternating optimization}

\newacronym{bcd}{BCD}{block coordinate descent}
\newacronym{bs}{BS}{base station}

\newacronym{csi}{CSI}{channel state information}

\newacronym{drl}{DRL}{deep reinforcement learning}

\newacronym{fcn}{FCN}{fully convolutional network}

\newacronym{iid}{i.i.d.}{independent and identically distributed}

\newacronym{los}{LoS}{line-of-sight}

\newacronym{mpc}{MPC}{multi-path component}
\newacronym{mrt}{MRT}{maximum ratio transmission}
\newacronym{mse}{MSE}{mean squared error}
\newacronym{mmse}{MMSE}{minimum mean squared error}
\newacronym{miso}{MISO}{multi-input-single-output}

\newacronym{nn}{NN}{neural network}
\newacronym{ris}{RIS}{reconfigurable intelligent surface}

\newacronym{snr}{SNR}{signal to noise ratio}
\newacronym{sdma}{SDMA}{space-division multiple access}

\newacronym{tsnr}{TSNR}{transmit signal to noise ratio}
\newacronym{wmmse}{WMMSE}{weighted minimum mean squared error}
\newacronym{wsr}{WSR}{weighted sum rate}

\newacronym{zf}{ZF}{zero forcing}
\setacronymstyle{long-short}
\glsdisablehyper

\DeclareMathOperator{\trace}{tr}
\newcommand*{\mat}[1]{\ensuremath{\mathbf{#1}}}

\usetikzlibrary{positioning}
\usetikzlibrary{shapes.geometric, arrows, decorations.pathreplacing}
\usetikzlibrary{shapes.arrows}
\usetikzlibrary{3d}
\pgfplotsset{compat=newest}

\newcommand{\mb}[1]{\mathbf{#1}}

\newcommand{\todo}[2][]{\ignorespaces
	\if\relax\detokenize{#1}\relax
	{\color{red}[TODO: #2]}%
	\else
	{\color{red}[TODO (#1): #2]}%
	\fi
}

\addto\extrasenglish{%

}
\usetikzlibrary{spy}

\begin{document}

\title{RISnet: A Scalable Approach for Reconfigurable Intelligent Surface Optimization with Partial CSI
\thanks{The work is supported in part by the Federal Ministry of Education and Research Germany (BMBF) as part of the 6G Research and Innovation Cluster 6G-RIC under Grant 16KISK031. The work of K.-L. Besser is supported by the German Research Foundation (DFG) under grant BE\,8098/1-1.
The work of V. Jamali is supported in part by the German Research Foundation (DFG) as part of project C8 within the Collaborative Research Center (CRC) 1053 -MAKI and in part by the LOEWE initiative (Hesse, Germany) within the emergenCITY center.}
}


\author{%
\IEEEauthorblockN{%
Bile Peng\IEEEauthorrefmark{1}, Karl-Ludwig Besser\IEEEauthorrefmark{2}, Ramprasad Raghunath\IEEEauthorrefmark{1}, Vahid Jamali\IEEEauthorrefmark{3}, Eduard A. Jorswieck\IEEEauthorrefmark{1}}
\IEEEauthorblockA{\IEEEauthorrefmark{1}Institute for Communications Technology, Technische Universit\"at Braunschweig, Germany}
\IEEEauthorblockA{\IEEEauthorrefmark{2}Department of Electrical and Computer Engineering, Princeton University, USA}
\IEEEauthorblockA{\IEEEauthorrefmark{3}Department of Electrical Engineering and Information Technology, Technische Universit\"at Darmstadt, Germany}
Email: \{b.peng, r.raghunath, e.jorswieck\}@tu-braunschweig.de, karl.besser@princeton.edu, vahid.jamali@tu-darmstadt.de
}

\maketitle

\begin{abstract}
The \gls{ris} is a promising technology that enables wireless communication systems to achieve improved performance by intelligently manipulating wireless channels.
In this paper, we consider the sum-rate maximization problem in a downlink multi-user \gls{miso} channel via \gls{sdma}.
Two major challenges of this problem are the high dimensionality due to the large number of \gls{ris} elements and the difficulty to obtain the full \gls{csi},
which is assumed known in many algorithms proposed in the literature.
Instead,
we propose a hybrid machine learning approach using the \gls{wmmse} precoder at the \gls{bs} and a dedicated \gls{nn} architecture, RISnet, for \gls{ris} configuration.
The RISnet has a good scalability to optimize 1296 \gls{ris} elements and requires partial \gls{csi} of only 16 \gls{ris} elements as input.
We show it achieves a high performance with low requirement for channel estimation for geometric channel models obtained with ray-tracing simulation.
The unsupervised learning lets the RISnet find an optimized \gls{ris} configuration by itself.
Numerical results show that a trained model configures the \gls{ris} with low computational effort, considerably outperforms the baselines,
and can work with discrete phase shifts.

\end{abstract}


\glsresetall

\section{Introduction}
The \gls{ris} is a promising technology that has been attracting increasing attention in recent years. 
An \gls{ris} is a planar array of many passive, reconfigurable elements that can reflect incoming electromagnetic waves with shifted complex phases~\cite{di2020smart}.
As an important application of the \gls{ris},
we consider the sum-rate maximization problem in a downlink \gls{ris}-assisted multi-user \gls{miso} channel with \gls{sdma},
which is a joint optimization problem of precoding at the \gls{bs} and the \gls{ris} configuration.
In the literature, the \gls{wmmse} precoder~\cite{shi2011iteratively} is proposed as an optimal precoder for weighted sum-rate maximization with \gls{sdma}.
On the \gls{ris} side, many different approaches for optimizing the \gls{ris} configuration have been presented in previous works.
This includes \gls{bcd} to maximize the weighted sum-rate~\cite{guo2020weighted}, majorization-maximization and the efficient alternating direction method of multipliers (ADMM) to optimize the sum-rate~\cite{huang2018achievable,liu2021two} and the sum-rate of multiple user groups~\cite{zhou2020intelligent}.
Riemannian manifold conjugate gradient (RMCG) and the Lagrangian method are applied to optimize multiple \glspl{ris} and \glspl{bs} to serve users on the cell edge~\cite{li2020weighted}.
Exhaustive search and successive refinement algorithm are applied to improve passive beamforming~\cite{wu2019beamforming}.
The configuration of active \glspl{ris} is optimized with the successive convex approximation algorithm to maximize the signal-to-noise ratio in~\cite{long2021active}.
Gradient-based optimization is applied to maximize the effective rank and the minimum singular value~\cite{Elmossallamy2021spatial}.

These analytical iterative methods do not scale well with the number of \gls{ris} elements in general.
No more than 100~\gls{ris} elements are assumed in \cite{guo2020weighted,huang2018achievable,zhou2020intelligent,liu2021two,li2020weighted,wu2019beamforming,long2021active}
and up to 400~\gls{ris} elements are assumed in \cite{Elmossallamy2021spatial},
which is still far from the vision of more than 1000~\gls{ris} elements~\cite{di2020smart}.
In fact, it has been shown that for most typical scenarios, 
large \glspl{ris} with several hundreds to several thousands of elements are needed in order to establish a sufficient link budget~\cite{najafi2020physics}.

Moreover, it is common that suboptimal approximations (e.g., \cite{guo2020weighted,wu2019beamforming}) have to be made for these approaches, which degrade the performance.
On the other hand,
machine learning models require much computational resources in training but the trained model can perform inference almost instantly.
They are more flexible due to the universal approximation theorem~\cite{hornik1989multilayer} such that they do not have to make suboptimal approximations.
If the model is trained properly on a representative training set,
the machine learning approach can achieve a better performance.
In recent years, deep learning~\cite{sheen2021deep,jiang2021learning,ozdougan2020deep}, deep reinforcement learning~\cite{feng2020deep} and meta learning~\cite{jung2021meta} have been applied to optimize the \gls{ris}.
However, scalability is still an open problem because the model complexity grows with the number of \gls{ris} elements.
Therefore, existing works employing machine learning only assume a limited number of \gls{ris} elements (no more than 100 in \cite{jiang2021learning,ozdougan2020deep,feng2020deep,jung2021meta} and 256 in \cite{sheen2021deep}).

Another common disadvantage of many references on \gls{ris}-assisted communication systems is the full \gls{csi} assumption (e.g., \cite{guo2020weighted,zhou2020intelligent,liu2021two,wu2019beamforming,long2021active,ozdougan2020deep}).
Due to the large number of \gls{ris} elements, the full \gls{csi} of all elements is extremely difficult to obtain in real time.
Possible contermeasures are, e.g., codebook-based \gls{ris} optimization~\cite{najafi2020physics,an2022codebook}.
However, the beam training is still a bottleneck.

To address the above issues, we introduce a \gls{nn} architecture \emph{RISnet}, which was first proposed in \cite{peng2023risnet} and receives significant improvements in this work, 
including the arbitrary number of users with permutation-invariance,
partial \gls{csi},
and validation with more realistic ray-tracing channel models.
The proposed approach has the following two advantages: 
\begin{enumerate}
    \item
We use the same filters to all \gls{ris} elements and users.
The number of parameters in the \gls{nn} is therefore independent from the number of \gls{ris} elements,
which improves the scalability significantly.
We demonstrate this by applying RISnet to a scenario with 4~users and 1296~\gls{ris} elements in \autoref{sec:results}.
    \item
RISnet supports partial \gls{csi} as input, which reduces the difficulty of channel estimation considerably (phase shifts of 1296~\gls{ris} elements are computed based on the \gls{csi} of 16 \gls{ris} elements).
\end{enumerate}
Furthermore, we show that our proposed approach computes the precoding matrix and \gls{ris} phase shifts in milliseconds, 
outperforms the baselines significantly,
and works with discrete phase shifts,
which is a more realistic assumption compared to continuous phase shifts for e.g., PIN-diode based \gls{ris} elements..

\section{System Model and Problem Formulation}
\begin{figure}[htbp]
    \centering
    \resizebox{.7\linewidth}{!}{
        \begin{tikzpicture}
        \tikzstyle{base}=[isosceles triangle, draw, rotate=90, fill=gray!60, minimum size =.5cm]
        \tikzstyle{user}=[rectangle, draw, rotate=90, fill=gray!60, minimum size =.5cm, rounded corners=0.1cm]
        \tikzstyle{element}=[rectangle, fill=gray!30]
        
        \node[base] (BS) at (-3,0){};
        \draw[decoration=expanding waves,decorate] (BS) -- (-3,1.7);
        \node[user] (UE1) at (4,2.5){};
        \node[user] (UE2) at (4,0){};
        \draw[step=0.33cm] (-1,2.95) grid (0, 3.98);
        \node (RIS) at (-0.5, 3.5) {};
        
        \node[below of=BS,yshift=.3cm]{BS};
        \node[right of=BS,xshift=-.3cm, yshift=.5cm]{$\mathbf{V}$};
        \node[below of=UE1,yshift=.3cm]{User 1};
        \node[below of=UE1,yshift=-.5cm]{\vdots};
        \node[below of=UE2,yshift=.3cm]{User $U$};
        \node[above of=RIS]{RIS};
        \node[above right=.15 and .3 of RIS]{$\boldsymbol{\Phi}$};

        \draw[-to,shorten >=3pt] (BS) to node[below=1mm, pos=.3] {$\mathbf{D}$} (UE1);
        \draw[-to,shorten >=3pt] (BS) to node[below=2mm] {} (UE2);
        
        \draw[-to] (BS) to node[above=4mm] {$\mathbf{H}$} (RIS);
        
        \draw[-to,shorten >=3pt] (RIS) to node[left=-1mm, below=0mm, pos=.3] {$\mathbf{G}$} (UE1);
        \draw[-to,shorten >=3pt] (RIS) to node[below=-3mm, left=2mm] {} (UE2);
    \end{tikzpicture}
    }
    \caption{System model.}
    \label{fig:system_model}
\end{figure}
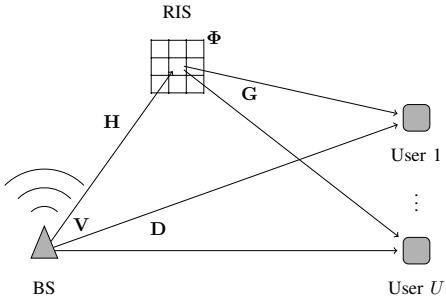

We consider a downlink \gls{ris} aided communication from a multi-antenna \gls{bs} to multiple users,
as shown in \autoref{fig:system_model}.
The channel from \gls{bs} to \gls{ris} is denoted as $\mathbf{H} \in \mathbb{C}^{N\times M}$,
where $N$ is the number of \gls{ris} elements
and $M$ is the number of \gls{bs} antennas.
The channel from \gls{ris} to users is denoted as $\mathbf{G} \in \mathbb{C}^{U\times N}$,
where $U$ is the number of users.
The channel from \gls{bs} directly to users is denoted as $\mathbf{D} \in \mathbb{C}^{U\times M}$.

The objective is to use \gls{sdma} to maximize the sum-rate of the users. 
The \gls{bs} can perform precoding subject to maximum transmit power constraint $E_{Tr}$.
Each \gls{ris} element can receive the signal from the \gls{bs}, shift the complex phase of the signal, and reflect signal without changing its amplitude.

We denote the precoding matrix as $\mb{V} \in \mathbb{C}^{M\times U}$.
The diagonal matrix $\boldsymbol{\Phi} \in \mathbb{C}^{N\times N}$ is the signal processing matrix at the \gls{ris}. The diagonal element in row $n$ and column $n$ is $\phi_{nn}=e^{j\psi_n}$, 
where $\psi_n \in [0, 2\pi)$ is the phase shift of \gls{ris} element~$n$. 
The signal received at the users is given as
\begin{equation}
\mathbf{y} = \left(\mathbf{G} \boldsymbol{\Phi} \mathbf{H} + \mathbf{D} \right) \mathbf{V} \mathbf{x} + \mathbf{n},
\label{eq:transmission_los}
\end{equation}
where $\mb{x} \in \mathbb{C}^{U \times 1}$ is the transmitted symbols,
$\mb{y} \in \mathbb{C}^{U \times 1}$ is the received symbols and $\mb{n} \in \mathbb{C}^{U \times 1}$ is the thermal noise. 

Let $\mathbf{C} \in \mathbb{C}^{U \times U}$ be the combined matrix of precoding and transmission, i.e.,
\begin{equation}
\mathbf{C} = (\mathbf{G} \boldsymbol{\Phi} \mathbf{H} + \mathbf{D}) \mathbf{V}
\label{eq:channel}
\end{equation}
and $c_{uv}$ be the element in row $u$ and column $v$ of $\mb{C}$.
The problem to maximize the sum-rate subject to the maximum transmit power can be given by
\begin{subequations}
\begin{align}
    \max_{\mathbf{V}, \boldsymbol{\Phi}}\quad & 
  L = \sum_{u=1}^U\log_2\left(1+\frac{|c_{uu}|^2}{\sum_{v\neq u}|c_{uv}|^2+\sigma^2}\right)
\\
    \text{s.t.}\quad & \trace{\left(\mathbf{V}\mathbf{V}^H\right)} \leq E_{Tr}\\
    & |\phi_{nn}|=1\\
    & |\phi_{nn'}|=0 \text{ for } n \neq n',
\end{align}
\label{eq:problem}
\end{subequations}
where $\sigma^2$ is the noise power.

In this problem, we optimize both precoding $\mathbf{V}$ at \gls{bs} and \gls{ris} phase shifts $\boldsymbol{\Phi}$.
While optimizing $\boldsymbol{\Phi}$ is a new and open problem,
the \gls{wmmse} precoder is proved to be the optimal precoder to maximize the \gls{wsr}.
It is shown that minimizing weighted \gls{mse} is equivalent to maximizing \gls{wsr}~\cite{shi2011iteratively}. 
By iteratively updating precoding vectors and weights of \gls{mse},
the \gls{wsr} is maximized.
We choose the \gls{wmmse} precoder for the precoding in \gls{bs}.
It is to note that $\mathbf{V}$ and $\boldsymbol{\Phi}$ must be jointly optimized because their optimal values depend on each other.
However, the iterative approach of \gls{wmmse} precoder makes it not differentiable and the gradient ascent cannot be applied to optimize the \gls{nn}.
To tackle this problem, we apply \gls{ao}, which updates $\mathbf{V}$ and phase shift matrices $\mathbf{\Phi}$ alternatively while keeping the other constant for all the data samples.

\section{Proposed Machine Learning Solution}
The aim of solving problem \eqref{eq:problem} is to jointly optimize the precoding matrix~$\mat{V}$ and the phase adjustments of the \gls{ris} elements~$\mat{\Phi}$.
Since it is difficult to find an optimal solution to this problem due to its high dimensionality (the large number of \gls{ris} elements), 
we propose solving \eqref{eq:problem} by employing machine learning.
In particular, we train an \gls{nn} to learn the mapping between the available \gls{csi}, the optimal precoding and phase adjustments.



\subsection{Framework of Unsupervised Machine Learning}
We define a neural network $N_\theta$
parameterized by $\theta$, which
maps from the \gls{csi} to the \gls{ris} phase shifts $\boldsymbol{\Phi}$.
Since $N_\theta$ cannot take the original \gls{csi} as input,
we define channel feature $\boldsymbol{\Gamma}$ as an equivalent \gls{csi} presentation, as will be explained in \autoref{sec:feature_definition}.
Hence, we have $\boldsymbol{\Phi} = N_\theta(\boldsymbol{\Gamma}).$
Applying the \gls{wmmse} precoder,
the objective function $L$ defined in \eqref{eq:problem}
is fully determined by $\boldsymbol{\Gamma}$
and $\boldsymbol{\Phi}$.
We can write the objective as $ L(\boldsymbol{\Gamma}, \boldsymbol{\Phi})=L(\boldsymbol{\Gamma}, N_\theta(\boldsymbol{\Gamma}); \theta).$
Note that we emphasize $L$ depends on the parameter $\theta$ given $\boldsymbol{\Gamma}$ here.
We collect massive channel features in a training set $\mathcal{D}$
and formulate the unsupervised machine learning problem as
\begin{equation}
    \max_\theta \sum_{\boldsymbol{\Gamma} \in \mathcal{D}} L(\boldsymbol{\Gamma}, N_\theta(\boldsymbol{\Gamma}); \theta).
    \label{eq:ml_objective}
\end{equation}
In this way, we optimize $N_\theta$ which maps from any $\boldsymbol{\Gamma} \in \mathcal{D}$ to $\boldsymbol{\Phi}$.
If the training set is general enough,
we would expect that a channel feature $\boldsymbol{\Gamma}' \notin \mathcal{D}$ can also be mapped to a good $\boldsymbol{\Phi}$.

\subsection{Channel Feature}
\label{sec:feature_definition}

In the \gls{ris}-assisted channel,
there are three channel matrices $\mathbf{H}$, $\mathbf{G}$ and $\mathbf{D}$,
among which
$\mathbf{H}$ is assumed to be constant because \gls{bs} and \gls{ris} are stationary and the environment is relatively invariant,
$\mathbf{G}$ and $\mathbf{D}$ depend on user positions and are inputs of $N_\theta$. 
We would like to define a feature $\boldsymbol{\gamma}_{un}$ for user~$u$ and \gls{ris} element $n$ such that we can apply the same filters to every user and \gls{ris} element to enable scalability.
Since $g_{un}$ in row~$u$ and column~$n$ of $\mathbf{G}$ is the channel gain from \gls{ris} element~$n$ to user~$u$,
we can simply include amplitude and phase of $g_{un}$ in $\boldsymbol{\gamma}_{un}$\footnote{The \gls{nn} does not take complex numbers as input.}.
On the other hand, elements in $\mathbf{D}$ cannot be mapped to \gls{ris} elements because $\mathbf{D}$ is the channel from \gls{bs} directly to users.
Therefore, we define $\mathbf{J}=\mathbf{D}\mathbf{H}^+$,
where $(\cdot)^+$ denotes the pseudo-inverse operation,
and \eqref{eq:transmission_los} becomes
$\mathbf{y} = \left(\mathbf{G} \boldsymbol{\Phi} + \mathbf{J}\right) \mathbf{H} \mathbf{V} \mathbf{x} + \mathbf{n}$,
i.e.,
signal $\mathbf{x}$ is precoded with $\mathbf{V}$, transmitted through channel $\mathbf{H}$ to the \gls{ris},
and through channel $\mathbf{G}\boldsymbol{\Phi} + \mathbf{J}$ to users. 
Elements $j_{un}$ of $\mathbf{J}$ is the channel gain from \gls{ris} element~$n$ to user~$u$.
The features of user~$u$ and \gls{ris} element~$n$ can then be defined as
$\boldsymbol{\gamma}_{un} = (|g_{un}|, \arg(g_{un}), |j_{un}|, \arg(j_{un}))^T \in \mathbb{R}^{4 \times 1}$,
where $|a|$ and $\arg(a)$ are amplitude and phase of complex number $a$, respectively.
The complete channel feature $\boldsymbol{\Gamma} \in \mathbb{R}^{4\times U \times N}$ is a three dimensional tensor,
where elements with index $u$ and $n$ in second and third dimensions are $\boldsymbol{\gamma}_{un}$.

\subsection{RISnet Architecture with Full CSI}
\label{sec:risnet_full}

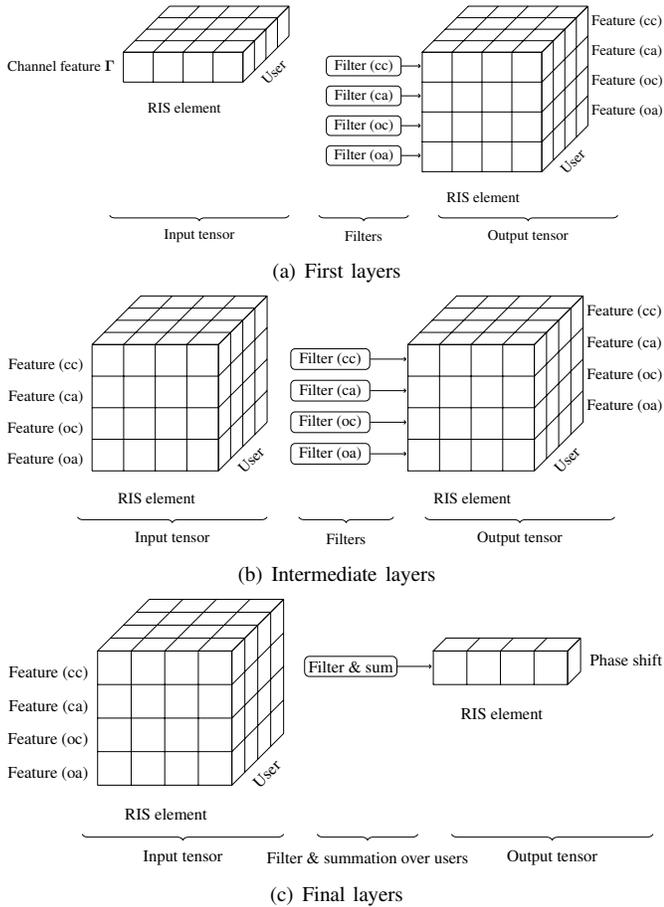
\begin{figure}[htbp]
    \centering
    \subfigure[First layers]{\resizebox{\linewidth}{!}{\begin{tikzpicture}
\makeatletter 
\tikzoption{canvas is xy plane at z}[]{%
  \def\tikz@plane@origin{\pgfpointxyz{0}{0}{#1}}%
  \def\tikz@plane@x{\pgfpointxyz{1}{0}{#1}}%
  \def\tikz@plane@y{\pgfpointxyz{0}{1}{#1}}%
  \tikz@canvas@is@plane
}
\makeatother
\NewDocumentCommand{\DrawCubes}{O {} m m m m m m}{%
    \def\XGridMin{#2}
    \def\XGridMax{#3}
    \def\YGridMin{#4}
    \def\YGridMax{#5}
    \def\ZGridMin{#6}
    \def\ZGridMax{#7}
    \begin{scope}[canvas is xy plane at z=\ZGridMax]
      \draw [#1] (\XGridMin,\YGridMin) grid (\XGridMax,\YGridMax);
    \end{scope}
    \begin{scope}[canvas is yz plane at x=\XGridMax]
      \draw [#1] (\YGridMin,\ZGridMin) grid (\YGridMax,\ZGridMax);
    \end{scope}
    \begin{scope}[canvas is xz plane at y=\YGridMax]
      \draw [#1] (\XGridMin,\ZGridMin) grid (\XGridMax,\ZGridMax);
    \end{scope}
}%

\tikzstyle{layer} = [rectangle, rounded corners, minimum width=2.5cm, minimum height=.6cm, align=center, text centered, draw=black]
    
y={(0.5cm,0.25cm)},x={(0.5cm,-0.25cm)},z={(0cm,{veclen(0.5,0.25)*1cm})}
    ]
    \DrawCubes [step=10mm,thin]{0}{4}{3}{4}{0}{4}
    \node (width) [rectangle, yshift=0.6cm, xshift=0.5cm,font=\Large] {RIS element};
    \node (height) [rectangle, yshift=2cm, xshift=-1.7cm, anchor=east,font=\Large] {Channel feature $\boldsymbol{\Gamma}$};
    \node (channels) [rectangle, yshift=1.8cm, xshift=3.5cm,rotate=45,font=\Large] {User};


\node (layer1) [layer, xshift=6.5cm, yshift=2.0cm,font=\Large] {Filter (cc)};
\node (layer2) [layer, below of=layer1, yshift=0cm,font=\Large] {Filter (ca)};
\node (layer3) [layer, below of=layer2, yshift=0cm,font=\Large] {Filter (oc)};
\node (layer4) [layer, below of=layer3, yshift=0cm,font=\Large] {Filter (oa)};

\draw [->] (layer1.east) -- (8.4, 2);
\draw [->] (layer2.east) -- (8.4, 1);
\draw [->] (layer3.east) -- (8.4, 0);
\draw [->] (layer4.east) -- (8.4, -1);

    \DrawCubes [step=10mm,thin]{9.99}{14}{0}{4}{0}{4}
    \node (width) [rectangle, yshift=-2.4cm, xshift=10.5cm,font=\Large] {RIS element};
    \node (height) [rectangle, yshift=3.5cm, xshift=14cm, anchor=west,font=\Large] {Feature (cc)};
    \node (height) [rectangle, yshift=2.5cm, xshift=14cm, anchor=west,font=\Large] {Feature (ca)};
    \node (height) [rectangle, yshift=1.5cm, xshift=14cm, anchor=west,font=\Large] {Feature (oc)};
    \node (height) [rectangle, yshift=0.5cm, xshift=14cm, anchor=west,font=\Large] {Feature (oa)};
    \node (channels) [rectangle, xshift=4cm, yshift=-1.2cm, xshift=9.5cm,rotate=45,font=\Large] {User};

\draw [decorate,decoration={brace,amplitude=5pt,mirror}]
  (-2,-3) -- (4,-3) node[midway, yshift=-.7cm, font=\Large]{Input tensor};
\draw [decorate,decoration={brace,amplitude=5pt,mirror}]
  (5,-3) -- (8,-3) node[midway, yshift=-.7cm, font=\Large]{Filters};
\draw [decorate,decoration={brace,amplitude=5pt,mirror}]
  (9,-3) -- (15,-3) node[midway, yshift=-.7cm, font=\Large]{Output tensor};
\end{tikzpicture}}}
    \subfigure[Intermediate layers]{\resizebox{\linewidth}{!}{\begin{tikzpicture}

\makeatletter 
\tikzoption{canvas is xy plane at z}[]{%
  \def\tikz@plane@origin{\pgfpointxyz{0}{0}{#1}}%
  \def\tikz@plane@x{\pgfpointxyz{1}{0}{#1}}%
  \def\tikz@plane@y{\pgfpointxyz{0}{1}{#1}}%
  \tikz@canvas@is@plane
}
\makeatother
\NewDocumentCommand{\DrawCubes}{O {} m m m m m m}{%
    \def\XGridMin{#2}
    \def\XGridMax{#3}
    \def\YGridMin{#4}
    \def\YGridMax{#5}
    \def\ZGridMin{#6}
    \def\ZGridMax{#7}
    \begin{scope}[canvas is xy plane at z=\ZGridMax]
      \draw [#1] (\XGridMin,\YGridMin) grid (\XGridMax,\YGridMax);
    \end{scope}
    \begin{scope}[canvas is yz plane at x=\XGridMax]
      \draw [#1] (\YGridMin,\ZGridMin) grid (\YGridMax,\ZGridMax);
    \end{scope}
    \begin{scope}[canvas is xz plane at y=\YGridMax]
      \draw [#1] (\XGridMin,\ZGridMin) grid (\XGridMax,\ZGridMax);
    \end{scope}
}%

\tikzstyle{layer} = [rectangle, rounded corners, minimum width=2.5cm, minimum height=.6cm, align=center, text centered, draw=black]
    
y={(0.5cm,0.25cm)},x={(0.5cm,-0.25cm)},z={(0cm,{veclen(0.5,0.25)*1cm})}
    ]
    \DrawCubes [step=10mm,thin]{0}{4}{0}{4}{0}{4}
    \node (width) [rectangle, yshift=-2.4cm, xshift=0.5cm,font=\Large] {RIS element};
    \node (height) [rectangle, yshift=1.8cm, xshift=-1.7cm, anchor=east,font=\Large] {Feature (cc)};
    \node (height) [rectangle, yshift=0.8cm, xshift=-1.7cm, anchor=east,font=\Large] {Feature (ca)};
    \node (height) [rectangle, yshift=-0.2cm, xshift=-1.7cm, anchor=east,font=\Large] {Feature (oc)};
    \node (height) [rectangle, yshift=-1.2cm, xshift=-1.7cm, anchor=east,font=\Large] {Feature (oa)};
    \node (channels) [rectangle, yshift=-1.2cm, xshift=3.5cm,rotate=45,font=\Large] {User};


\node (layer1) [layer, xshift=6cm, yshift=2.0cm,font=\Large] {Filter (cc)};
\node (layer2) [layer, below of=layer1, yshift=0cm,font=\Large] {Filter (ca)};
\node (layer3) [layer, below of=layer2, yshift=0cm,font=\Large] {Filter (oc)};
\node (layer4) [layer, below of=layer3, yshift=0cm,font=\Large] {Filter (oa)};

\draw [->] (layer1.east) -- (8.4, 2);
\draw [->] (layer2.east) -- (8.4, 1);
\draw [->] (layer3.east) -- (8.4, 0);
\draw [->] (layer4.east) -- (8.4, -1);

    \DrawCubes [step=10mm,thin]{9.99}{14}{0}{4}{0}{4}
    \node (width) [rectangle, yshift=-2.4cm, xshift=10.5cm,font=\Large] {RIS element};
    \node (height) [rectangle, yshift=3.5cm, xshift=14cm, anchor=west,font=\Large] {Feature (cc)};
    \node (height) [rectangle, yshift=2.5cm, xshift=14cm, anchor=west,font=\Large] {Feature (ca)};
    \node (height) [rectangle, yshift=1.5cm, xshift=14cm, anchor=west,font=\Large] {Feature (oc)};
    \node (height) [rectangle, yshift=0.5cm, xshift=14cm, anchor=west,font=\Large] {Feature (oa)};
    \node (channels) [rectangle, xshift=4cm, yshift=-1.2cm, xshift=9.5cm,rotate=45,font=\Large] {User};

\draw [decorate,decoration={brace,amplitude=5pt,mirror}]
  (-2,-3) -- (4,-3) node[midway, yshift=-.7cm, font=\Large]{Input tensor};
\draw [decorate,decoration={brace,amplitude=5pt,mirror}]
  (5,-3) -- (8,-3) node[midway, yshift=-.7cm, font=\Large]{Filters};
\draw [decorate,decoration={brace,amplitude=5pt,mirror}]
  (9,-3) -- (15,-3) node[midway, yshift=-.7cm, font=\Large]{Output tensor};
\end{tikzpicture}}}
    \subfigure[Final layers]{\resizebox{\linewidth}{!}{\begin{tikzpicture}

\makeatletter 
\tikzoption{canvas is xy plane at z}[]{%
  \def\tikz@plane@origin{\pgfpointxyz{0}{0}{#1}}%
  \def\tikz@plane@x{\pgfpointxyz{1}{0}{#1}}%
  \def\tikz@plane@y{\pgfpointxyz{0}{1}{#1}}%
  \tikz@canvas@is@plane
}
\makeatother
\NewDocumentCommand{\DrawCubes}{O {} m m m m m m}{%
    \def\XGridMin{#2}
    \def\XGridMax{#3}
    \def\YGridMin{#4}
    \def\YGridMax{#5}
    \def\ZGridMin{#6}
    \def\ZGridMax{#7}
    \begin{scope}[canvas is xy plane at z=\ZGridMax]
      \draw [#1] (\XGridMin,\YGridMin) grid (\XGridMax,\YGridMax);
    \end{scope}
    \begin{scope}[canvas is yz plane at x=\XGridMax]
      \draw [#1] (\YGridMin,\ZGridMin) grid (\YGridMax,\ZGridMax);
    \end{scope}
    \begin{scope}[canvas is xz plane at y=\YGridMax]
      \draw [#1] (\XGridMin,\ZGridMin) grid (\XGridMax,\ZGridMax);
    \end{scope}
}%

\tikzstyle{layer} = [rectangle, rounded corners, minimum width=2.5cm, minimum height=.6cm, align=center, text centered, draw=black]
    
y={(0.5cm,0.25cm)},x={(0.5cm,-0.25cm)},z={(0cm,{veclen(0.5,0.25)*1cm})}
    ]
    \DrawCubes [step=10mm,thin]{0}{4}{0}{4}{0}{4}
    \node (width) [rectangle, yshift=-2.4cm, xshift=0.5cm,font=\Large] {RIS element};
    \node (height) [rectangle, yshift=1.8cm, xshift=-1.7cm, anchor=east,font=\Large] {Feature (cc)};
    \node (height) [rectangle, yshift=0.8cm, xshift=-1.7cm, anchor=east,font=\Large] {Feature (ca)};
    \node (height) [rectangle, yshift=-0.2cm, xshift=-1.7cm, anchor=east,font=\Large] {Feature (oc)};
    \node (height) [rectangle, yshift=-1.2cm, xshift=-1.7cm, anchor=east,font=\Large] {Feature (oa)};
    \node (channels) [rectangle, yshift=-1.2cm, xshift=3.5cm,rotate=45,font=\Large] {User};

\node (layer1) [layer, xshift=6cm, yshift=2.0cm,font=\Large] {Filter \& sum};

\draw [->] (layer1.east) -- (8.4, 2);

    \DrawCubes [step=10mm,thin]{9.99}{14}{3}{4}{3}{4}
    \node (width) [rectangle, yshift=0.6cm, xshift=10.5cm,font=\Large] {RIS element};
    \node (height) [rectangle, yshift=2.2cm, xshift=13cm, anchor=west,font=\Large] {Phase shift};

\draw [decorate,decoration={brace,amplitude=5pt,mirror}]
  (-2,-3) -- (4,-3) node[midway, yshift=-.7cm, font=\Large]{Input tensor};
\draw [decorate,decoration={brace,amplitude=5pt,mirror}]
  (5,-3) -- (8,-3) node[midway, yshift=-.7cm, font=\Large]{Filter \& summation over users};
\draw [decorate,decoration={brace,amplitude=5pt,mirror}]
  (9,-3) -- (15,-3) node[midway, yshift=-.7cm, font=\Large]{Output tensor};
\end{tikzpicture}}}
    \caption{Information processing of layers in RISnet.}
    \label{fig:info_processing}
\end{figure}

The RISnet comprises $L$ layers.
In each layer,
we would like to apply the same filters to all users and \gls{ris} elements
in order to realize scalability.
However,
the optimal phase shift of an \gls{ris} element depends on all users and all \gls{ris} elements.
Therefore, in every layer of the \gls{nn},
we apply four filters for information processing of 
current user and current \gls{ris} element (cc),
current user and all \gls{ris} elements (ca),
other users and current \gls{ris} element (oc), and
other users and all \gls{ris} elements (oa).
Denote the input feature of user~$u$ and \gls{ris} element~$n$ in layer~$i$ as $\mathbf{f}_{un,i}$
(in particular, $\mathbf{f}_{un, 1} = \boldsymbol{\gamma}_{un}$),
the output feature of user~$u$ and \gls{ris} element~$n$ in layer~$i$ is calculated as
\begin{equation}
\begin{aligned}
&\mathbf{f}_{un, i + 1}\\ = &
\left(
\begin{array}{c}
     \text{ReLU}(\mathbf{W}^{cc}_{i} \mathbf{f}_{un, i} + \mathbf{b}_i^{cc}) \\
     \left(\sum_{n'}\text{ReLU}(\mathbf{W}^{ca}_{i} \mathbf{f}_{un', i} + \mathbf{b}_i^{ca})\right) \big/ N\\
     \left(\sum_{u'\neq u}\text{ReLU}(\mathbf{W}^{oc}_{i} \mathbf{f}_{u'n, i} + \mathbf{b}_i^{oc})\right) \big/ (U-1)\\
     \left(\sum_{u'\neq u}\sum_{n'}\text{ReLU}(\mathbf{W}^{oa}_{i} \mathbf{f}_{u'n', i} + \mathbf{b}_i^{oa})\right) \big/ \\ \hspace{12em} (N(U-1))\\
\end{array}
\right)
\end{aligned}
\label{eq:layer_processing}
\end{equation}
for $i < L$,
where 
$\mathbf{W}^{cc}_i \in \mathbb{R}^{Q_i \times P_i}$ is trainable weights of class cc in layer $i$ with the input feature dimension $P_i$ in layer~$i$ (i.e., $\mathbf{f}_{un, i} \in \mathbb{R}^{P_i \times 1}$) and output feature dimension $Q_i$ in layer~$i$ of class cc,
$\mathbf{b}^{cc}_i \in \mathbb{R}^{Q_i \times 1}$ is trainable bias of class cc in layer $i$.
Similar definitions and same dimensions apply to classes ca, oc and oa.
For class cc in layer~$i$,
the output feature of user~$u$ and \gls{ris} element~$n$ is computed by applying a conventional full-connected layer (a linear transform with weights $\mathbf{W}^{cc}_i$ and bias $\mathbf{b}^{cc}_i$ and the ReLU activation) to $\mathbf{f}_{un,i}$ (the first line of \eqref{eq:layer_processing}).
For class ca in layer~$i$,
we first apply the conventional linear transform with weights $\mathbf{W}^{ca}_i$ and bias $\mathbf{b}^{ca}_i$ and ReLU activation to $\mathbf{f}_{un,i}$,
then compute the mean value of all \gls{ris} elements.
Therefore, the output feature of class ca for user~$u$ and all \gls{ris} elements is the same (the second line of \eqref{eq:layer_processing}).
For classes oc and oa,
the information processing is similar but the output feature of user~$u$ is not computed with the input feature of user~$u$,
but is averaged over results of users other than user~$u$ (the third and fourth lines of \eqref{eq:layer_processing}).
We can infer from the above description that $\mathbf{f}_{un, i + 1} \in \mathbb{R}^{4Q_i \times 1}$ for all $u$ and $n$
because the output feature comprises of four classes.
Therefore $P_{i + 1} = 4 Q_i$.
The whole output feature
$\mathbf{F}_{i + 1} \in \mathbb{R}^{4Q_i \times U \times N}$ is a three dimensional tensor,
where elements with index $u$ and $n$ in second and third dimensions are $\mathbf{f}_{un, i + 1}$.
We see from \eqref{eq:layer_processing} that all four parts of $\mathbf{f}_{un,i+1}$ use $\mathbf{F}_i$ to compute the output features.
The output feature of the current user and \gls{ris} element depends exclusively on the current user and \gls{ris} element (class cc, i.e., first part of $\mathbf{f}_{un, i + 1}$) and
the output feature of the other users and/or all \gls{ris} elements is the mean of the raw output of other users and/or all \gls{ris} elements (other classes, i.e., second to fourth parts of $\mathbf{f}_{un, i + 1}$).
Therefore, it should contain sufficient local and global information to make a wise decision on $\psi_n$.
For the final layer,
we use one filter to process the information.
Features of different users are summed up to be the phase shifts because all the users share the same phase shift.
Because the information processing before the final layer is symmetric to users
and addition in the final layer is commutative,
the RISnet is permutation-invariant to users,
i.e., if we permute the users in the input,
the output of \gls{ris} phase shifts does not change.
This is a desired property that can enhance the generality of the \gls{nn}.
The information processing of a layer is illustrated in \autoref{fig:info_processing}.

In this way,
we achieve a good scalability with the RISnet.
However,
it is difficult to obtain the full \gls{csi} ($\boldsymbol{\gamma}_{un}$ for every user and \gls{ris} element) in practice.
Therefore, we propose to use partial \gls{csi} as input to RISnet. 

\subsection{RISnet Architecture with Partial CSI}

In this section,
we assume an \gls{ris} architecture where only a few \gls{ris} elements are equipped with RF chains for channel estimation.
Only a partial \gls{csi} can be estimated with pilot signals from the users by these \gls{ris} elements.
If the propagation paths in the channel are mostly specular or \gls{los} paths that contribute to channel gains of all \gls{ris} elements,
the 
the full \gls{csi} can be inferred from partial \gls{csi} because all \gls{ris} elements share the same propagation paths (with different path loss and complex phase).
This fact suggests that we can use partial \gls{csi} as input to RISnet and compute phase shifts for all \gls{ris} elements.
Rather than performing an explicit full \gls{csi} prediction like an image super-resolution,
we perform an implicit full \gls{csi} prediction,
i.e., an end-to-end learning from partial \gls{csi} to full \gls{ris} configuration.

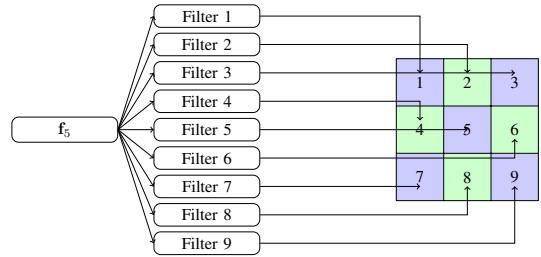
\begin{figure}[htbp]
    \centering
    \resizebox{.8\linewidth}{!}{    \begin{tikzpicture}
\tikzstyle{block} = [rectangle, rounded corners, text width=2cm, text centered, draw=black]
    
\node (feature) [block] {$\mathbf{f}_5$};
\node (filter1) [block, right of=feature, xshift=2cm, yshift=2.4cm] {Filter 1};
\node (filter2) [block, below of=filter1, yshift=.4cm] {Filter 2};
\node (filter3) [block, below of=filter2, yshift=.4cm] {Filter 3};
\node (filter4) [block, below of=filter3, yshift=.4cm] {Filter 4};
\node (filter5) [block, below of=filter4, yshift=.4cm] {Filter 5};
\node (filter6) [block, below of=filter5, yshift=.4cm] {Filter 6};
\node (filter7) [block, below of=filter6, yshift=.4cm] {Filter 7};
\node (filter8) [block, below of=filter7, yshift=.4cm] {Filter 8};
\node (filter9) [block, below of=filter8, yshift=.4cm] {Filter 9};

\fill[blue!20!white, xshift=7cm, yshift=-1.5cm] (0, 0) rectangle (1, 1);
\fill[green!20!white, xshift=7cm, yshift=-1.5cm] (0, 1) rectangle (1, 2);
\fill[blue!20!white, xshift=7cm, yshift=-1.5cm] (0, 2) rectangle (1, 3);
\fill[green!20!white, xshift=7cm, yshift=-1.5cm] (1, 0) rectangle (2, 1);
\fill[blue!20!white, xshift=7cm, yshift=-1.5cm] (1, 1) rectangle (2, 2);
\fill[green!20!white, xshift=7cm, yshift=-1.5cm] (1, 2) rectangle (2, 3);
\fill[blue!20!white, xshift=7cm, yshift=-1.5cm] (2, 0) rectangle (3, 1);
\fill[green!20!white, xshift=7cm, yshift=-1.5cm] (2, 1) rectangle (3, 2);
\fill[blue!20!white, xshift=7cm, yshift=-1.5cm] (2, 2) rectangle (3, 3);

\draw[step=1cm, yshift=.5cm] (7,-2) grid (10, 1);

\draw[-to] (feature.east) -- (filter1.west);
\draw[-to] (feature.east) -- (filter2.west);
\draw[-to] (feature.east) -- (filter3.west);
\draw[-to] (feature.east) -- (filter4.west);
\draw[-to] (feature.east) -- (filter5.west);
\draw[-to] (feature.east) -- (filter6.west);
\draw[-to] (feature.east) -- (filter7.west);
\draw[-to] (feature.east) -- (filter8.west);
\draw[-to] (feature.east) -- (filter9.west);

\draw[-to] (filter1.east) -| (7.5, 1.2);
\draw[-to] (filter2.east) -| (8.5, 1.2);
\draw[-to] (filter3.east) -- (9.5, 1.2);
\draw[-to] (filter4.east) -| (7.5, 0.2);
\draw[-to] (filter5.east) -- (8.5, 0);
\draw[-to] (filter6.east) -| (9.5, -0.2);
\draw[-to] (filter7.east) -- (7.5, -1.2);
\draw[-to] (filter8.east) -| (8.5, -1.2);
\draw[-to] (filter9.east) -| (9.5, -1.2);

\node at (7.5, 1) {1};
\node at (8.5, 1) {2};
\node at (9.5, 1) {3};
\node at (7.5, 0) {4};
\node at (8.5, 0) {5};
\node at (9.5, 0) {6};
\node at (7.5, -1) {7};
\node at (8.5, -1) {8};
\node at (9.5, -1) {9};
\end{tikzpicture}}
    \caption{Application of 9 filters to expand from one anchor RIS element to 9 RIS elements,
    where $\mathbf{f}_5$ is the channel feature of \gls{ris} element 5.
    Indices of user and layer are omitted for simplicity since the expansion is for \gls{ris} elements.}
    \label{fig:expansion_filter}
\end{figure}

The input channel feature for the RISnet with partial \gls{csi} in this section is channel features of all users and \gls{ris} elements with the ability to estimate the \gls{csi}.
In the first layer,
only these \gls{ris} elements are taken into consideration.
Define an \gls{ris} element which has been taken into consideration of the RISnet as an \emph{anchor element},
and a layer in RISnet that expands the anchor elements as an \emph{expansion layer},
the RISnet expands the anchor elements from the \gls{ris} elements with \gls{csi} to all \gls{ris} elements.
The basic idea of the expansion layer is to apply the same filter to an adjacent \gls{ris} element with the same relative position to the anchor element,
as shown in \autoref{fig:expansion_filter}.
Since one anchor element is expanded to 9 anchor elements in an expansion layer,
there are 9 filters for each class (cc, ca, oc and oa).
Instead of applying one filter to every \gls{ris} element of a class as output of itself, 
as described in \autoref{sec:risnet_full},
we apply 9 filters to every anchor element as output of the same \gls{ris} element and the adjacent 8 \gls{ris} elements.
The output of \gls{ris} element~$n$ using filter~$j$ is computed as
\begin{equation}
\begin{aligned}
&\mathbf{f}_{u\nu(n, j), i + 1}\\ = &
\left(
\begin{array}{c}
     \text{ReLU}(\mathbf{W}^{cc}_{i,j} \mathbf{f}_{un, i} + \mathbf{b}_{i,j}^{cc}) \\
     \left(\sum_{n'}\text{ReLU}(\mathbf{W}^{co}_{i,j} \mathbf{f}_{un', i} + \mathbf{b}_{i,j}^{co})\right) \big/ N\\
     \left(\sum_{u'\neq u}\text{ReLU}(\mathbf{W}^{oc}_{i,j} \mathbf{f}_{u'n, i} + \mathbf{b}_{i,j}^{oc})\right) \big/ (U-1)\\
     \left(\sum_{u'\neq u}\sum_{n'}\text{ReLU}(\mathbf{W}^{oo}_{i,j} \mathbf{f}_{u'n', i} + \mathbf{b}_{i,j}^{oo})\right) \big/ \\ \hspace{12em} (N(U-1))\\
\end{array}
\right),
\end{aligned}
\label{eq:expansion_layer_processing}
\end{equation}
where $\nu(n, j)$ is the \gls{ris} element index when applying filter~$j$ for input of \gls{ris} element~$n$.
According to \autoref{fig:expansion_filter} and assuming that the \gls{ris} element index begins from 1 with the upper left corner,
increases first along rows and then changes to the next row (i.e., the index in row~$w$ and column~$h$ is $h + (w - 1)\cdot H$, where $H$ is the number of columns of the \gls{ris} array.),
we have
\begin{equation}
    \nu(n, j)= 
\begin{cases}
n - H - 2 + j & j = 1, 2, 3,\\
n - 5 + j & j = 4, 5, 6,\\
n + H - 8 + j & j = 7, 8, 9.\\
\end{cases}
\label{eq:nu}
\end{equation}

\begin{figure}
    \centering
    \subfigure[First expansion\label{fig:expansion-partial1}]{\resizebox{.6\linewidth}{!}{    \begin{tikzpicture}

\fill[blue!20!white] (0, 0) rectangle (3, 3);
\fill[green!20!white] (0, 3) rectangle (3, 6);
\fill[blue!20!white] (0, 6) rectangle (3, 9);
\fill[green!20!white] (3, 0) rectangle (6, 3);
\fill[blue!20!white] (3, 3) rectangle (6, 6);
\fill[green!20!white] (3, 6) rectangle (6, 9);
\fill[blue!20!white] (6, 0) rectangle (9, 3);
\fill[green!20!white] (6, 3) rectangle (9, 6);
\fill[blue!20!white] (6, 6) rectangle (9, 9);

\fill[blue!40!white] (1, 1) rectangle (2, 2);
\fill[blue!40!white] (4, 1) rectangle (5, 2);
\fill[blue!40!white] (7, 1) rectangle (8, 2);
\fill[blue!40!white] (1, 4) rectangle (2, 5);
\fill[blue!40!white] (7, 4) rectangle (8, 5);
\fill[blue!40!white] (1, 7) rectangle (2, 8);
\fill[blue!40!white] (4, 7) rectangle (5, 8);
\fill[blue!40!white] (7, 7) rectangle (8, 8);

\foreach \i in {4, 13, 22, 31}
\foreach \j in {4, 13, 22, 31}
{
\fill[blue] (\i, \j) rectangle (\i+1, \j+1);
}
\draw[step=1cm, gray!50!white] (0,0) grid (36, 36);
\end{tikzpicture}}}
    \subfigure[Second expansion\label{fig:expansion-partial2}]{\resizebox{.6\linewidth}{!}{    \begin{tikzpicture}

\fill[blue!20!white] (0, 0) rectangle (1, 1);
\fill[green!20!white] (0, 1) rectangle (1, 2);
\fill[blue!20!white] (0, 2) rectangle (1, 3);
\fill[green!20!white] (1, 0) rectangle (2, 1);
\fill[blue!20!white] (1, 1) rectangle (2, 2);
\fill[green!20!white] (1, 2) rectangle (2, 3);
\fill[blue!20!white] (2, 0) rectangle (3, 1);
\fill[green!20!white] (2, 1) rectangle (3, 2);
\fill[blue!20!white] (2, 2) rectangle (3, 3);

\foreach \i in {1, 4, 7, 10, 13, 16, 19, 22, 25, 28, 31, 34}
\foreach \j in {1, 4, 7, 10, 13, 16, 19, 22, 25, 28, 31, 34}
{
\fill[blue] (\i, \j) rectangle (\i+1, \j+1);
}

\foreach \i in {4, 13, 22, 31}
\foreach \j in {4, 13, 22, 31}
{
\fill[blue] (\i, \j) rectangle (\i+1, \j+1);
}
\draw[step=1cm, gray!50!white] (0,0) grid (36, 36);
\end{tikzpicture}}}
    \caption{Expansion of considered RIS elements. Blue: anchor RIS elements.
    Lower left corner: example of the expansion to extend the anchor RIS elements from the blue element to the adjacent elements (light blue elements in Subfigure~(a) and all elements in Subfigure~(b)).}
    \label{fig:expansion}
\end{figure}
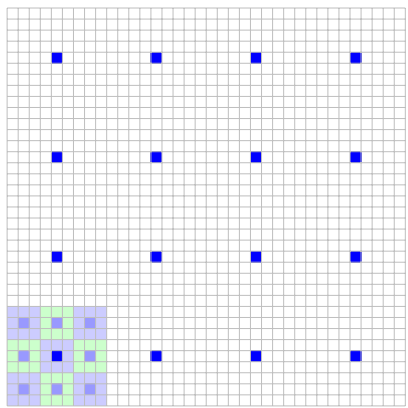
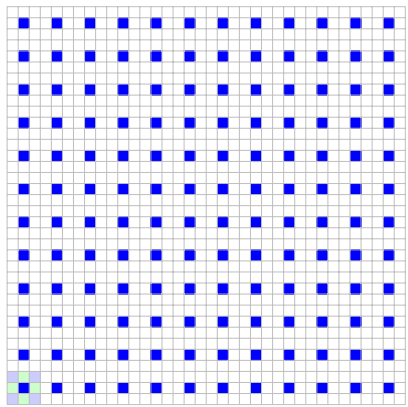

By defining two such expansion layers,
the numbers of anchor elements are increased by a factor of 9 in both row and column.
If we have 16 \gls{ris} elements (4$\times$4) with \gls{csi},
we can generate phase shifts of 1296 (36$\times$36) \gls{ris} elements\footnote{Note that the factor of 9 is fixed given the expansion layer structure and the number of expansion layers.
The number of total \gls{ris} elements is determined according to the performance requirement.
Besides, the number of \gls{ris} elements with \gls{csi} is required to be large enough to infer the full \gls{csi}.
The assumption on number of \gls{ris} elements with \gls{csi} and the \gls{ris} size is made based on the above considerations.}.
This process is illustrated in \autoref{fig:expansion},
where the blue \gls{ris} elements in \autoref{fig:expansion-partial1} can estimate the channel from the pilot signals from the users.
Such \gls{ris} elements are only 1/81 of all \gls{ris} elements.
In an RISnet of 8 layers,
layer 3 and 6 are such expansion layers,
such that the expansion layers are roughly equally placed among the 8 layers.
Other layers are normal layers described in \autoref{sec:risnet_full}.
It is to note that we can use tensor multiplication to implement \eqref{eq:layer_processing}, \eqref{eq:expansion_layer_processing}
and use permutation matrix to sort the \gls{ris} elements in the correct order according to \eqref{eq:nu},
such that we can utilize the power of parallel computing to accelerate training and testing.

During the model training,
the full \gls{csi} is still required to compute the sum-rate with \eqref{eq:problem}.
However, phase shifts of all \gls{ris} elements $\boldsymbol{\Phi}$ are computed with the partial \gls{csi},
which means that only the partial \gls{csi} is required in application.

The training process is formulated in \autoref{alg}.

\begin{algorithm}[H]
\caption{RISnet training}
\label{alg}
\begin{algorithmic}[1]
\State Randomly initialize RISnet.
\Repeat
\State Randomly select a batch of data samples.
\State Compute precoding matrix with current \gls{ris} configuration.
\State Compute $\boldsymbol{\Phi}=N_\theta(\boldsymbol{\Gamma})$ for every data sample.
\State Compute the channel $\mathbf{C}$ for every data sample.
\State Compute the objective function for every data sample.
\State Compute the gradient of the objective w.r.t. the neural network parameters
\State Perform a stochastic gradient ascent step
\Until{Predefined number of iterations achieved}
\end{algorithmic}
\end{algorithm}
\section{Training and Testing Results}
\label{sec:results}

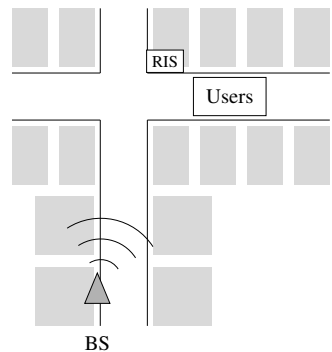
\begin{figure}
    \centering
    \resizebox{.5\linewidth}{!}{
    		\begin{tikzpicture}[scale=1]
            \tikzstyle{base}=[isosceles triangle, draw, rotate=90, fill=gray!60, minimum size =0.12cm]
   
			\foreach \i in {1, 1.8, 3.4, 4.2, 5.0, 5.8}
			\foreach \j in {2, 4}
			{
				\fill[gray!30!white] (\i, \j) rectangle (\i+0.6, \j+1);
			}
   
			\foreach \i in {1.4, 3.4}
			\foreach \j in {-0.4, 0.8}
			{
				\fill[gray!30!white] (\i, \j) rectangle (\i+1, \j+1);
			}
			\node[align=center, rectangle,draw, minimum height=.65cm, text width=1cm] (ue) at (4.7,3.5) {Users};
			\draw (1,3.9) -- (2.5,3.9);
			\draw (1,3.1) -- (2.5,3.1);
			\draw (3.3,3.9) -- (6.4,3.9);
			\draw (3.3,3.1) -- (6.4,3.1);
			
			\draw (2.5,-0.4) -- (2.5,3.1);
			\draw (3.3,-0.4) -- (3.3,3.1);
			\draw (2.5,3.9) -- (2.5,5);
			\draw (3.3,3.9) -- (3.3,5);
			
            \node[base] (BS) at (2.45,0.1){};
			\node[below of=BS, yshift=.2cm] (bs) {BS};
            \draw[decoration=expanding waves,decorate] (BS) -- (3.2,1.9, 1.2);
			\node[rectangle, fill=white, draw, scale=.8] (ris) at (3.6,4.1)
			{  \hspace{.0cm}RIS\hspace{.0cm} };
	\end{tikzpicture}}
    \caption{The ray-tracing scenario.}
    \label{fig:raytracing}
\end{figure}

We use the open-source DeepMIMO ray-tracing data set~\cite{Alkhateeb2019} to generate realistic ray-tracing channel data,
where positions of \gls{bs} and \gls{ris} are fixed and positions of users are uniformly distributed in the given area.
We generate a training set and a testing set with different random user positions,
i.e., the channel data in training set and testing set are different, and independently and identically distributed.
The positions of \gls{bs}, \gls{ris} and users are chosen to satisfy the following criterion:
1) There are \gls{los} propagation paths between \gls{bs} and \gls{ris}, \gls{ris} and users,
but not between \gls{bs} and users directly,
otherwise the direct channel is so strong that the \gls{ris} does not play a significant row.
2) There are \glspl{mpc} between \gls{bs} and \gls{ris} such that the channel matrix $\mathbf{H}$ has a high rank
because the number of served users cannot be greater than the matrix rank.
3) The minimum distance between users in a group for \gls{sdma} is 8~meters,
such that the channel matrix $\mathbf{G}$ has a high rank.
Following these criteria, the designed scenario is shown in \autoref{fig:raytracing}.

We choose the Adam optimizer with a learning rate of $10^{-3}$ for the training.
A batch with 512 data samples is randomly sampled for a training iteration.
We compute the \gls{wmmse} precoding matrix $\mathbf{V}$ with the current \gls{ris} configuration for each data sample.
The gradient of \eqref{eq:problem} is computed with respect to the neural network parameter $\theta$,
which is updated with the Adam optimizer.

\begin{figure}[htbp]
\subfigure[Deterministic ray-tracing model\label{fig:training-determ-ray-trace}]{\input{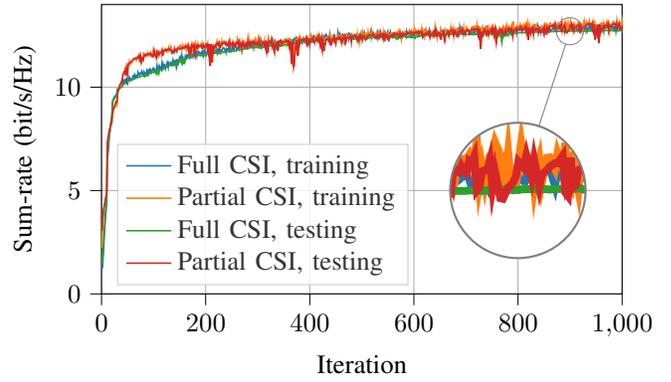}}
\subfigure[Deterministic ray-tracing model plus \gls{iid} complex Gaussian gain\label{fig:training-determ-ray-trace-plus-gauss}]{\input{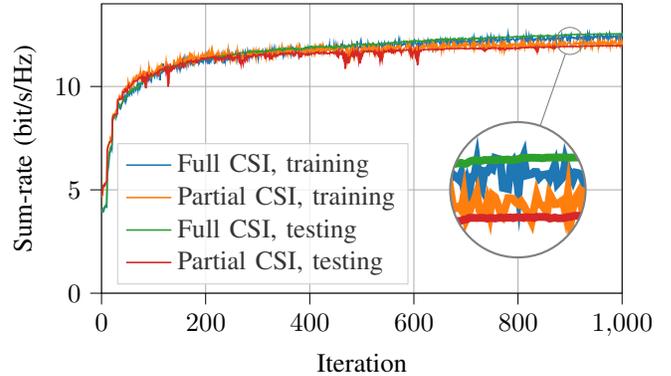}}
\subfigure[\Gls{iid} complex Gaussian channel gains\label{fig:training-complex-gauss}]{\input{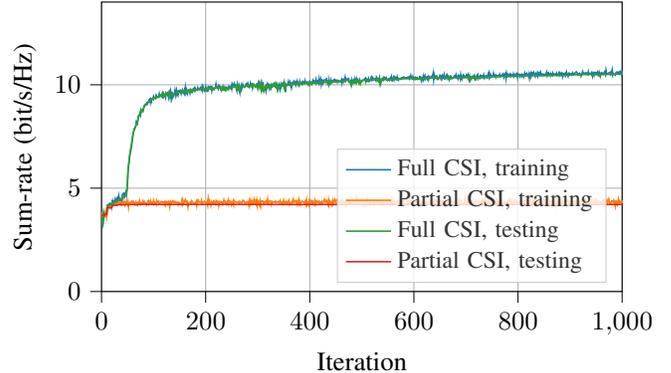}}
\caption{Training results with different spatial correlation between RIS elements.}
\label{fig:training}
\end{figure}

In the ray-tracing model,
all the \gls{ris} elements share the same propagation paths
(e.g., \gls{los} path, reflection path on the ground or on a building)
with different pathlosses and complex phases due to the different positions of the \gls{ris} element.
The strongest path (the \gls{los} path) is at least 1.5 times stronger than the second-strongest path.
There are up to four propagation paths from the \gls{ris} to a user due to the limited number of reflectors nearby.
In this case,
the \gls{csi} of a few \gls{ris} elements would be representative of \gls{csi} of all \gls{ris} elements.
In the opposite extreme case,
if the channel comprises infinitely many and weak \glspl{mpc} due to isotropic scattering,
channel gains are \gls{iid} complex Gaussian random variables in the \gls{ris} elements~\cite{jamali2022impact}.
In this case, the full \gls{csi} cannot be inferred from the partial \gls{csi}.
Between the two extreme cases,
the channel gain is contributed by both deterministic strong \glspl{mpc} and \gls{iid} complex Gaussian gain due to weak scattering.
Training and testing results using the above three channel models are shown in \autoref{fig:training}.

Besides the observation that the training and testing results are very similar,
which suggests there is no overfitting,
an important conclusion is that the performance with partial \gls{csi} depends on the channel model.
Assuming the ray-tracing channel model
(i.e., there are multiple specular propagation paths, which contribute to channel gains of all \gls{ris} elements),
RISnet with partial \gls{csi} achieves a similar performance to the RISnet with full \gls{csi} (\autoref{fig:training-determ-ray-trace}).
The ray-tracing channel model fulfills the requirement to apply the partial \gls{csi}.
On the contrary,
assuming \gls{iid} complex Gaussian channel gain,
the 1/81 partial \gls{csi} does not provide sufficient information about the full channel
and RISnet with with partial \gls{csi} does not work (\autoref{fig:training-complex-gauss}).
Between the two extreme cases,
\gls{iid} channel gains due to infinitely many and weak scattering \glspl{mpc} exist but are not dominant compared to the deterministic strong propagation paths.
RISnet with partial \gls{csi} has a slightly worse performance than RISnet with full \gls{csi} because the channel gain is contributed by both deterministic paths and \gls{iid} scattering paths (\autoref{fig:training-determ-ray-trace-plus-gauss}).

The contribution of scattering paths depends mainly on the number, sizes and roughness of the scatterers as well as the frequency.
The more, rougher and smaller the scatterers are in the surrounding environment,
the more important scattering paths there are in the channel~\cite{ju2019scattering}.
According to~\cite{rappaport2002wireless},
scattering can be safely ignored in frequency band between 300\,MHz and 3\,GHz,
suggesting that the proposed method would work in this frequency band.

\begin{figure}[htbp]
    \centering
\begin{tikzpicture}

\definecolor{crimson2143940}{RGB}{214,39,40}
\definecolor{darkgray176}{RGB}{176,176,176}
\definecolor{darkorange25512714}{RGB}{255,127,14}
\definecolor{forestgreen4416044}{RGB}{44,160,44}
\definecolor{lightgray204}{RGB}{204,204,204}
\definecolor{mediumpurple148103189}{RGB}{148,103,189}
\definecolor{sienna1408675}{RGB}{140,86,75}
\definecolor{steelblue31119180}{RGB}{31,119,180}

\begin{axis}[
legend cell align={left},
legend style={
  fill opacity=0.8,
  draw opacity=1,
  text opacity=1,
  at={(0.03,-1.1)},
  anchor=south west,
  draw=lightgray204
},
tick align=outside,
tick pos=left,
x grid style={darkgray176},
xmin=-0.571428571428571, xmax=2.57142857142857,
xtick style={color=black},
xtick={0,1,2},
xticklabels={Det.,Det. + i.i.d. gain,I.i.d.},
y grid style={darkgray176},
ylabel={Sum-rate (bit/s/Hz)},
ymin=0, ymax=13.5114,
ytick style={color=black},
width=\linewidth,
height=.2\textheight
]
\draw[draw=none,fill=steelblue31119180] (axis cs:-0.428571428571429,0) rectangle (axis cs:-0.285714285714286,12.768);
\addlegendimage{ybar,ybar legend,draw=none,fill=steelblue31119180}
\addlegendentry{Full CSI, continuous phase shifts}

\draw[draw=none,fill=steelblue31119180] (axis cs:0.571428571428571,0) rectangle (axis cs:0.714285714285714,12.56);
\draw[draw=none,fill=steelblue31119180] (axis cs:1.57142857142857,0) rectangle (axis cs:1.71428571428571,10.528);
\draw[draw=none,fill=darkorange25512714] (axis cs:-0.285714285714286,0) rectangle (axis cs:-0.142857142857143,12.268);
\addlegendimage{ybar,ybar legend,draw=none,fill=darkorange25512714}
\addlegendentry{Full CSI, discrete phase shifts}

\draw[draw=none,fill=darkorange25512714] (axis cs:0.714285714285714,0) rectangle (axis cs:0.857142857142857,11.952);
\draw[draw=none,fill=darkorange25512714] (axis cs:1.71428571428571,0) rectangle (axis cs:1.85714285714286,10.104);
\draw[draw=none,fill=forestgreen4416044] (axis cs:-0.142857142857143,0) rectangle (axis cs:0,12.868);
\addlegendimage{ybar,ybar legend,draw=none,fill=forestgreen4416044}
\addlegendentry{Partial CSI, continuous phase shifts}

\draw[draw=none,fill=forestgreen4416044] (axis cs:0.857142857142857,0) rectangle (axis cs:1,11.96);
\draw[draw=none,fill=forestgreen4416044] (axis cs:1.85714285714286,0) rectangle (axis cs:2,4.212);
\draw[draw=none,fill=crimson2143940] (axis cs:0,0) rectangle (axis cs:0.142857142857143,11.508);
\addlegendimage{ybar,ybar legend,draw=none,fill=crimson2143940}
\addlegendentry{Partial CSI, discrete phase shifts}

\draw[draw=none,fill=crimson2143940] (axis cs:1,0) rectangle (axis cs:1.14285714285714,10.352);
\draw[draw=none,fill=crimson2143940] (axis cs:2,0) rectangle (axis cs:2.14285714285714,4.044);

\draw[draw=none,fill=mediumpurple148103189] (axis cs:0.142857142857143,0) rectangle (axis cs:0.285714285714286,1.528);
\addlegendimage{ybar,ybar legend,draw=none,fill=mediumpurple148103189}
\addlegendentry{Random continuous phase shifts}
\draw[draw=none,fill=mediumpurple148103189] (axis cs:1.14285714285714,0) rectangle (axis cs:1.28571428571429,1.591);
\draw[draw=none,fill=mediumpurple148103189] (axis cs:2.14285714285714,0) rectangle (axis cs:2.28571428571429,1.304);

\draw[draw=none,fill=sienna1408675] (axis cs:0.285714285714286,0) rectangle (axis cs:0.428571428571429,9.219);
\addlegendimage{ybar,ybar legend,draw=none,fill=sienna1408675}
\addlegendentry{BCD, continuous phase shifts~\cite{guo2020weighted}}
\draw[draw=none,fill=sienna1408675] (axis cs:1.28571428571429,0) rectangle (axis cs:1.42857142857143,9.11);
\draw[draw=none,fill=sienna1408675] (axis cs:2.28571428571429,0) rectangle (axis cs:2.42857142857143,8.902);
\end{axis}

\end{tikzpicture}
    \caption{Test results with different spatial correlation between RIS elements and settings (Det. = deterministic ray-tracing model, I.i.d. = \gls{iid} complex Gaussian channel gain).}
    \label{fig:test}
\end{figure}
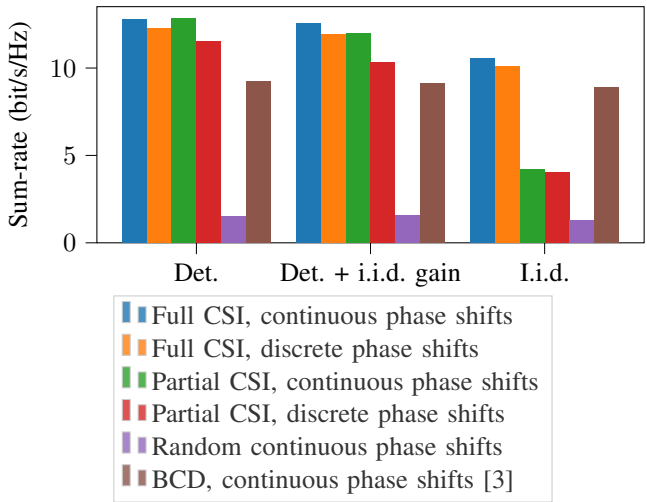

\autoref{fig:test} shows the achieved sum-rate with three channel models and different settings,
where the discrete phase shifts are 0, $\frac{1}{2}\pi$, $\pi$ and $\frac{3}{2}\pi$.
Simple rounding is applied to convert continuous phase shift to discrete phase shift.
We choose random phase shift and the \gls{bcd} algorithm~\cite{guo2020weighted} as baselines.
We observe that the difference between continuous and discrete phase shifts is only marginal for the resolution of $\frac{1}{2}\pi$.
Sum-rate with partial \gls{csi} is comparable to sum-rate with full \gls{csi} for deterministic ray-tracing channel model and deterministic channel model plus \gls{iid} perturbation.
Moreover, the RISnet outperforms the two baselines considerably except RISnet with partial \gls{csi} and \gls{iid} channel models.
The \gls{bcd} algorithm employs full \gls{csi} but applies suboptimal approximation,
which makes it more efficient but the performance is degraded.
Nevertheless, it still requires more than 10 minutes to run 2000 iterations in order to compute the suboptimal \gls{ris} phase shifts for one channel realization.
Compared to it, although the training of RISnet takes a long time,
application of a trained RISnet to a new channel realization is done in a few milliseconds,
and the performance is better than the \gls{bcd} algorithm,
which makes the proposed method more suitable as a high-performance, real-time solution.

\section{Conclusion}

The \gls{ris} is a promising technology for future wireless communications to optimize the channel property for better performance,
but it faces two major challenges: 
scalability and the assumption of full \gls{csi}, which can be difficult to obtain in practical scenarios.
In this work,
we propose an \gls{nn} architecture RISnet,
which is scalable because its number of parameters is independent from the number of \gls{ris} elements
(we assume 1296 \gls{ris} elements in the simulation)
and requires \gls{csi} of a small portion of elements (16 elements in our simulation, i.e., 1/81 of all elements).
The RISnet and the \gls{wmmse} precoder are applied for \gls{sdma} for a multi-antenna \gls{bs} and multiple users.
Simulation results show that the RISnet with full \gls{csi} outperforms two baselines for all types of channels
and the RISnet with partial \gls{csi} outperforms the baselines for deterministic ray-tracing channels and ray-tracing channels plus \gls{iid} complex Gaussian channel gains due to scattering.
Moreover, discrete phase shifts with a granularity of $\frac{1}{2}\pi$ only cause moderate performance loss.
Source code and data of this paper are available under \url{https://github.com/bilepeng/risnet_partial_csi}.

\printbibliography

@inproceedings{jung2021meta,
  title={Meta-learning for 6{G} communication networks with reconfigurable intelligent surfaces},
  author={Jung, Minchae and Saad, Walid},
  booktitle={ICASSP 2021-2021 IEEE International Conference on Acoustics, Speech and Signal Processing (ICASSP)},
  pages={8082--8086},
  year={2021},
  organization={IEEE}
}

@article{sheen2021deep,
  title={A deep learning based modeling of reconfigurable intelligent surface assisted wireless communications for phase shift configuration},
  author={Sheen, Baoling and Yang, Jin and Feng, Xianglong and Chowdhury, Md Moin Uddin},
  journal={IEEE Open Journal of the Communications Society},
  volume={2},
  pages={262--272},
  year={2021},
  publisher={IEEE}
}

@article{wu2019beamforming,
  title={Beamforming optimization for wireless network aided by intelligent reflecting surface with discrete phase shifts},
  author={Wu, Qingqing and Zhang, Rui},
  journal={IEEE Transactions on Communications},
  volume={68},
  number={3},
  pages={1838--1851},
  year={2019},
  publisher={IEEE}
}

@article{di2020smart,
  title={Smart radio environments empowered by reconfigurable intelligent surfaces: How it works, state of research, and the road ahead},
  author={Di Renzo, Marco and Zappone, Alessio and Debbah, Merouane and Alouini, Mohamed-Slim and Yuen, Chau and De Rosny, Julien and Tretyakov, Sergei},
  journal={IEEE Journal on Selected Areas in Communications},
  volume={38},
  number={11},
  pages={2450--2525},
  year={2020},
  publisher={IEEE}
}

@article{jiang2021learning,
  title={Learning to reflect and to beamform for intelligent reflecting surface with implicit channel estimation},
  author={Jiang, Tao and Cheng, Hei Victor and Yu, Wei},
  journal={IEEE Journal on Selected Areas in Communications},
  volume={39},
  number={7},
  pages={1931--1945},
  year={2021},
  publisher={IEEE}
}

@inproceedings{ozdougan2020deep,
  title={Deep learning-based phase reconfiguration for intelligent reflecting surfaces},
  author={{\"O}zdo{\u{g}}an, {\"O}zgecan and Bj{\"o}rnson, Emil},
  booktitle={2020 54th Asilomar Conference on Signals, Systems, and Computers},
  pages={707--711},
  year={2020},
  organization={IEEE}
}

@InProceedings{Alkhateeb2019,
author = {Alkhateeb, A.},
title = {{DeepMIMO}: A Generic Deep Learning Dataset for Millimeter Wave and Massive {MIMO} Applications},
booktitle = {Proc. of Information Theory and Applications Workshop (ITA)},
year = {2019},
pages = {1-8},
Address = {San Diego, CA}, }

@article{an2022codebook,
  title={Codebook-based solutions for reconfigurable intelligent surfaces and their open challenges},
  author={An, Jiancheng and Xu, Chao and Wu, Qingqing and Ng, Derrick Wing Kwan and Di Renzo, Marco and Yuen, Chau and Hanzo, Lajos},
  journal={IEEE Wireless Communications},
  year={2022},
  publisher={IEEE}
}

@inproceedings{ju2019scattering,
  title={Scattering mechanisms and modeling for terahertz wireless communications},
  author={Ju, Shihao and Shah, Syed Hashim Ali and Javed, Muhammad Affan and Li, Jun and Palteru, Girish and Robin, Jyotish and Xing, Yunchou and Kanhere, Ojas and Rappaport, Theodore S},
  booktitle={ICC 2019-2019 IEEE International Conference on Communications (ICC)},
  pages={1--7},
  year={2019},
  organization={IEEE}
}

@article{rappaport2002wireless,
  title={Wireless Communications--Principles and Practice},
  author={Rappaport, Theodore S},
  journal={Microwave Journal},
  volume={45},
  number={12},
  pages={128--129},
  year={2002},
  publisher={Horizon House Publications, Inc.}
}

@article{shi2011iteratively,
  title={An iteratively weighted {MMSE} approach to distributed sum-utility maximization for a {MIMO} interfering broadcast channel},
  author={Shi, Qingjiang and Razaviyayn, Meisam and Luo, Zhi-Quan and He, Chen},
  journal={IEEE Transactions on Signal Processing},
  volume={59},
  number={9},
  pages={4331--4340},
  year={2011},
  publisher={IEEE}
}

@article{zhou2020intelligent,
  title={Intelligent reflecting surface aided multigroup multicast {MISO} communication systems},
  author={Zhou, Gui and Pan, Cunhua and Ren, Hong and Wang, Kezhi and Nallanathan, Arumugam},
  journal={IEEE Transactions on Signal Processing},
  volume={68},
  pages={3236--3251},
  year={2020},
  publisher={IEEE}
}

@article{guo2020weighted,
  title={Weighted sum-rate maximization for reconfigurable intelligent surface aided wireless networks},
  author={Guo, Huayan and Liang, Ying-Chang and Chen, Jie and Larsson, Erik G},
  journal={IEEE Transactions on Wireless Communications},
  volume={19},
  number={5},
  pages={3064--3076},
  year={2020},
  publisher={IEEE}
}

@inproceedings{liu2021two,
  title={Two-user {SINR} Region for Reconfigurable Intelligent Surface Aided Downlink Channel},
  author={Liu, Xian and Sun, Cong and Jorswieck, Eduard A},
  booktitle={2021 IEEE International Conference on Communications Workshops (ICC Workshops)},
  pages={1--6},
  year={2021},
  organization={IEEE}
}

@inproceedings{huang2018achievable,
  title={Achievable rate maximization by passive intelligent mirrors},
  author={Huang, Chongwen and Zappone, Alessio and Debbah, M{\'e}rouane and Yuen, Chau},
  booktitle={2018 IEEE International Conference on Acoustics, Speech and Signal Processing (ICASSP)},
  pages={3714--3718},
  year={2018},
  organization={IEEE}
}

@article{li2020weighted,
  title={Weighted sum-rate maximization for multi-{IRS} aided cooperative transmission},
  author={Li, Zhengfeng and Hua, Meng and Wang, Qingxia and Song, Qingheng},
  journal={IEEE Wireless Communications Letters},
  volume={9},
  number={10},
  pages={1620--1624},
  year={2020},
  publisher={IEEE}
}

@article{Elmossallamy2021spatial,
archivePrefix = {arXiv},
arxivId = {2009.07064},
author = {Elmossallamy, Mohamed A. and Zhang, Hongliang and Sultan, Radwa and Seddik, Karim G. and Song, Lingyang and Han, Zhu and Han, Zhu},
doi = {10.1109/LWC.2020.3025030},
eprint = {2009.07064},
primaryclass = {eess.SP},
journal = {IEEE Wireless Communications Letters},
number = {2},
pages = {226--230},
title = {{On Spatial Multiplexing Using Reconfigurable Intelligent Surfaces}},
volume = {10},
year = {2021}
}

@article{feng2020deep,
  title={Deep reinforcement learning based intelligent reflecting surface optimization for {MISO} communication systems},
  author={Feng, Keming and Wang, Qisheng and Li, Xiao and Wen, Chao-Kai},
  journal={IEEE Wireless Communications Letters},
  volume={9},
  number={5},
  pages={745--749},
  year={2020},
  publisher={IEEE}
}

@inproceedings{peng2023risnet,
  title={{RISnet}: a Dedicated Scalable Neural Network Architecture for Optimization of Reconfigurable Intelligent Surfaces},
  author={Peng, Bile and Siegismund-Poschmann, Finn and Jorswieck, Eduard A},
  booktitle={WSA \& SCC 2023; 26th International ITG Workshop on Smart Antennas and 13th Conference on Systems, Communications, and Coding},
  pages={1--6},
  year={2023},
  organization={VDE}
}

@article{hornik1989multilayer,
  title={Multilayer feedforward networks are universal approximators},
  author={Hornik, Kurt and Stinchcombe, Maxwell and White, Halbert},
  journal={Neural networks},
  volume={2},
  number={5},
  pages={359--366},
  year={1989},
  publisher={Elsevier}
}

@article{long2021active,
  title={Active reconfigurable intelligent surface-aided wireless communications},
  author={Long, Ruizhe and Liang, Ying-Chang and Pei, Yiyang and Larsson, Erik G},
  journal={IEEE Transactions on Wireless Communications},
  volume={20},
  number={8},
  pages={4962--4975},
  year={2021},
  publisher={IEEE}
}

@article{najafi2020physics,
  title={Physics-based modeling and scalable optimization of large intelligent reflecting surfaces},
  author={Najafi, Marzieh and Jamali, Vahid and Schober, Robert and Poor, H Vincent},
  journal={IEEE Transactions on Communications},
  volume={69},
  number={4},
  pages={2673--2691},
  year={2020},
  publisher={IEEE}
}

@inproceedings{jamali2022impact,
  title={Impact of Channel Models on Performance Characterization of {RIS}-Assisted Wireless Systems},
  author={Jamali, Vahid and Ghanem, Walid and Schober, Robert and Poor, H Vincent},
  booktitle={European Conference on Antennas and Propagation (EuCAP)},
  year={2023}
}
\end{document}